\documentclass[12pt,preprint]{aastex}

\shorttitle{Early B-type supergiants in NGC 300}
\shortauthors{Urbaneja et al.}

\begin{document}

\title{Quantitative spectral analysis of early B-type supergiants in the
Sculptor galaxy NGC 300\footnote{Based on observations obtained at the ESO Very Large Telescope}}

%

\author{Miguel Alejandro Urbaneja, Artemio Herrero\footnote{Instituto de Astrof\'{\i}sica de Canarias,
V\'{\i}a L\'actea S/N, E-38200 La Laguna, Canary Islands, Spain, maup@ll.iac.es, ahd@ll.iac.es; Dpto. de Astrof\'{\i}sica, Universidad de 
La Laguna, Avda. Astrof\'{\i}sico Francisco Sanch\'ez, E-38271 La Laguna, 
Canary Islands, Spain, ahd@ll.iac.es}~, Fabio Bresolin, Rolf-Peter Kudritzki\footnote{Institute for Astronomy, 
University of Hawaii, 2680 Woodlawn Drive, Honolulu, Hawaii 96822, bresolin@ifa.hawaii.edu, kud@ifa.hawaii.edu}~, 
Wolfgang Gieren\footnote{Universidad de Concepci\'on, Departamento de F\'{\i}sica, Casilla 160-C, 
Concepci\'on, Chile, wgieren@coma.cfm.udec.cl}~ and Joachim Puls\footnote{Universit\"ats-Sternwarte M\"unchen, 
Scheinerstr. 1, D-81679 M\"unchen, Germany, uh101aw@usm.uni-muenchen.de}}

\begin{abstract}

The spectra of two early B-type supergiant stars in the Sculptor spiral galaxy
NGC 300 are analysed by means of non-LTE line blanketed unified model
atmospheres, aimed at determining their chemical composition and the fundamental
stellar and wind parameters. For the first time a detailed
chemical abundance pattern (He, C, N, O, Mg and Si) is obtained for a B-type
supergiant beyond the Local Group. The derived stellar properties are consistent 
with those of other Local Group B-type 
supergiants of similar types and metallicities. One of the stars shows a near solar 
metallicity while the other one resembles more a SMC B supergiant. The effects of the
lower metallicity can be detected in the derived wind momentum.

\end{abstract}


\section{Introduction}

The 8-10 meter class telescopes and their new generation instruments make it
possible to extend the quantitative stellar spectroscopy beyond the Local Group. Early 
B-type supergiant
stars are ideal targets for detailed spectroscopy even at low resolution
(R$\sim$1000). Their blue spectra are rich in metal features which allows us the
analysis of chemical species like C, N, O, Si and Mg. Although our knowledge of
the evolution of massive stars still has open questions, most of the recent works 
indicate that the blue luminous supergiants do not show any contamination
of their oxygen surface abundances during the early stages of their evolution, neither the
O-types \citep[][]{villamariz2002}, nor the B-types \citep[][]{smartt1997, monteverde2000,  
smartt2002}, nor the A-types \citep[][]{venn1995, takeda1998, przybilla2002}, which enables
a direct comparison between the stellar oxygen abundances and the ones derived from 
\ion{H}{2} regions. 
This has become extremely important, especially in the extragalactic field where oxygen
is used as the primary metallicity indicator, due to
the fact that 
at high metallicity (larger than approx. 0.5 solar) strong line methods must be used,
for which the choice of the calibration strongly influences the derived abundances
\citep[][]{kewley2002, pilyugin2002}.
In addition to chemical abundance studies, blue
luminous stars have strong radiatively driven mass outflows which can provide 
us with information on extragalatic distances by means of the Wind Momentum -
Luminosity Relationship, WLR \citep[][and references therein]{kudritzki2000}.

Recently, and within a wide program aimed at the spectroscopy study of luminous
blue stars beyond the Local Group, first steps have been done for
A-type supergiants in NGC 3621 \citep[6.7 Mpc away,][]{bresolin2001}.
Quantitative spectroscopy has been shown to be possible for A-type supergiants
\citep[][]{bresolin2002a} and Wolf-Rayet stars \citep[][]{bresolin2002b} 
in NGC 300, 2.02 Mpc away in the Sculptor group. Here we report the first 
quantitative analysis of B-type
supergiants (hereafter B-Sg) out of the Local Group, presenting the detailed 
chemical pattern along with the stellar parameters and the wind properties. The 
technique will be applied in a forthcoming paper to a large set of early B-Sg 
located at several
galactocentric distances in order to derive radial abundance gradients of the
$\alpha$-elements. Combined with the results of a similar study
of A-type supergiants it will provide a wealth of information on the chemical
evolution of the host galaxy NGC 300.

\section{Observations}

The stars are part of a spectroscopic survey of photometrically selected blue
luminous supergiants in the Sculptor galaxy NGC 300, obtained at the VLT with
the FORS multiobject spectrograph, and described in detail by
\citet[][]{bresolin2002a}, which presents a spectral catalog of 70 luminous blue
supergiants in the blue region ($\sim$ 4000 - 5000 \AA). The selected stars are
identified as B-12 and A-9 in that spectral catalog (see their Table 2 and finding
charts). In September 2001 
the spectra of the H$\alpha$ region were obtained in order to measure the mass-loss
rates, which provide us with a complete coverage of
the 3800 - 7200 \AA~wavelength range at R$\sim$1000 resolution. The 
reader is referred to
\citet[][]{bresolin2002a} for a detailed description of the observations and
reduction process, as well as for the photometry and the spectral classification 
of the stars.

\section{Spectral analysis}

The spectra of early B-Sg are
dominated by the \ion{O}{2} lines, followed by
\ion{N}{2}/\ion{N}{3}, \ion{Si}{3}/\ion{Si}{4}, \ion{C}{2}/\ion{C}{3} and \ion{Mg}{2}, 
in addition to H and \ion{He}{1} lines. At high resolution it is 
possible to detect some other metal lines of \ion{Al}{3}, \ion{S}{2}/\ion{S}{3} and 
\ion{Fe}{3} but, due to their intrinsic weakness, these lines do not have any influence 
in the analysis at low resolution and could hardly be used to fix the abundance of such 
elements. Fig. \ref{fig1} shows the high resolution - high S/N ratio (R$\sim$15000, 
SNR$\sim$350) blue
spectrum of the Galactic supergiant HD14956 (B1.5Ia), and the same spectrum degraded to the
resolution of the NGC 300 data, R$\sim$1000 (labeled as {\it \#d} in the figure). We have 
also included the identification of the more important
lines. As can be 
seen, only a few strong lines remain isolated at that low resolution, therefore the
analysis must be based on the comparison of the observed spectra to a set of
model atmospheres that include a vast number of lines in the calculation of the
emergent fluxes. We have taken into account more than two hundreds metal lines
in the 3800 - 6000 \AA~wavelength range. It is important to include extense metal 
line lists because of the fact
that some spectral features are formed by the contribution of several
chemical species (e.g. the strong blend of O, N and C at $\sim$ 4650 \AA).
We have excluded some strong isolated lines because our atomic models do
not consider the levels involved in these transitions.  
Nevertheless, these lines are isolated and have no influence on the results.

Even considering the noise effects in the lower resolution FORS spectra (displayed also 
in Fig. \ref{fig1}), strong metal features can still be 
detected and used for a detailed chemical abundance analysis. In particular a wealth of 
information can be extracted from 
the selected regions at 4070, 4320, 4420 (\ion{O}{2}), 4550 - 4570 (\ion{Si}{3}),
4600 - 4660 (\ion{O}{2}, \ion{N}{2}, \ion{N}{3} and \ion{C}{3}) and 5010 (\ion{N}{2} 
and \ion{He}{1}).

\subsection{Atmosphere models}

We use the newest version of the FASTWIND code \citep[first presented by][]{santolayarey1997} which 
solves the radiation transfer in a moving
media by means of suitable approximations which simplify the numerical treatment of the problem but
without affecting the physical significance of the results. The atmospheric structure is treated in a
consistent way, assuming a $\beta$-velocity law in the wind, ensuring a smooth transition between the
"photosphere" and the "wind"; the temperature structure is approximated by means of {\it non-LTE Hopf
functions} carefully chosen to ensure the flux conservation better than 2 \% at any depth
point; rate equations are solved in the co-moving frame scheme, with the coupling between the radiation
field and the rate equations solved using local ALOs \citep[following][]{puls1991}. This new version
includes the effects of the {\it line blanketing}. The reader is referred to Puls et al. 
(2003, in preparation) for a detailed description. We have analysed two Galactic stars, 10 Lac (O9V) 
and HD209975 (O9.5Ib) in order to compare our results with the ones obtained with other codes. In the case 
of 10 Lac, our results agree with the recent ones by \citet[][see their comparison to the results by 
Hubeny et al. 1998]{herrero2002}. The derived parameters for HD209975 are consistent with the results by
\citet[][]{villamariz2002} which used plane-parallel model with line blocking.

A model is prescribed by the effective temperature $T_{eff}$, the surface gravity {\it log g}, the
stellar radius $R_*$ (all these three quantities are defined at $\tau_{Ross.} = 2/3$), the 
mass-loss rate $\dot{M}$, the wind terminal velocity
$v_\infty$, the $\beta$ exponent of the wind velocity law, the He abundance 
$Y_{He}$, the microturbulent velocity $v_{turb}$ and, in the case of B-type stars, the {\it Si} abundance.
The $T_{eff}$ is well determined from the \ion{Si}{3} triplet and the blends of \ion{Si}{4} (with 
\ion{O}{2}
at 4090 \AA~and with \ion{O}{2}/\ion{He}{1} at 4120 \AA), and the surface gravity from the Balmer
hydrogen lines, provided that the mass-loss rate information is extracted from the H$\alpha$ profile. An
important issue concerns the wind terminal velocity, that must be adopted from a
spectral type - v$_\infty$ empirical calibration \citep[][]{haser1995, kudritzki2000}. The assumed
terminal velocity affects the derived $\dot{M}$ and the {\it log g}. But, with the joined information
from H$\alpha$ and H$\beta$, the mass-loss rate and v$_\infty$ can be constrained to yield reasonable
uncertainties in {\it log g}. 
The stellar radius is derived interactively from the absolute magnitude,
deduced from the apparent magnitude after adopting a distance modulus \citep[$\mu =
26.53$, ][]{freedman2001}, and the
%
%
model emergent flux \citep[][]{kudritzki1999}, which also provides the reddening by the
comparison of the synthetic colors with the observed ones.
 
\subsection{Results}

\begin{figure}
 \plotone{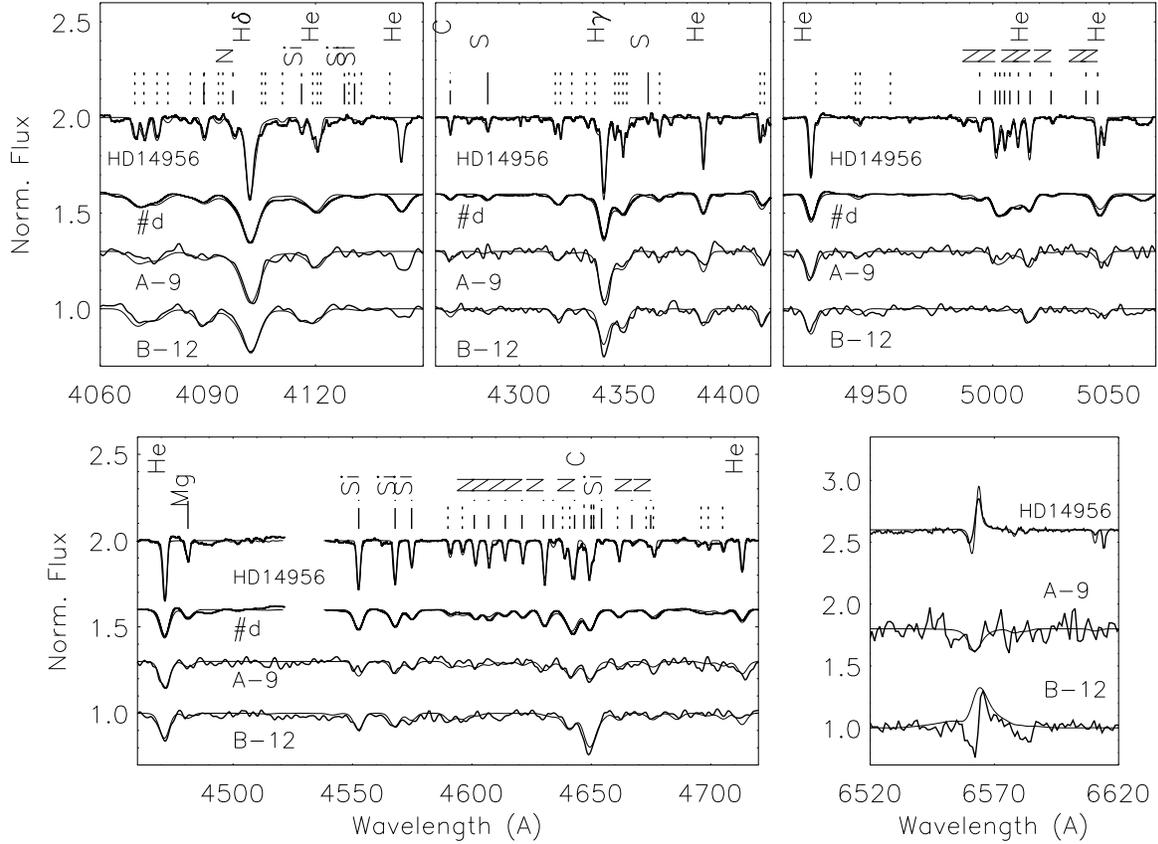}
 \caption{Observed spectra of B-12 and A-9, the Galactic 
 B-type supergiant HD14956, the same spectrum degraded to the NGC 300 data resolution and the 
 final fits. The spectra have been shifted for the sake of clarity. An identification of most 
 prominent features are given: O (dotted, without labels), Si (dashed-dotted-dotted), 
 N (dashed), C (dashed-dotted-dashed), Mg (solid), H and He. \label{fig1}}
\end{figure}

\begin{figure}
 \plotone{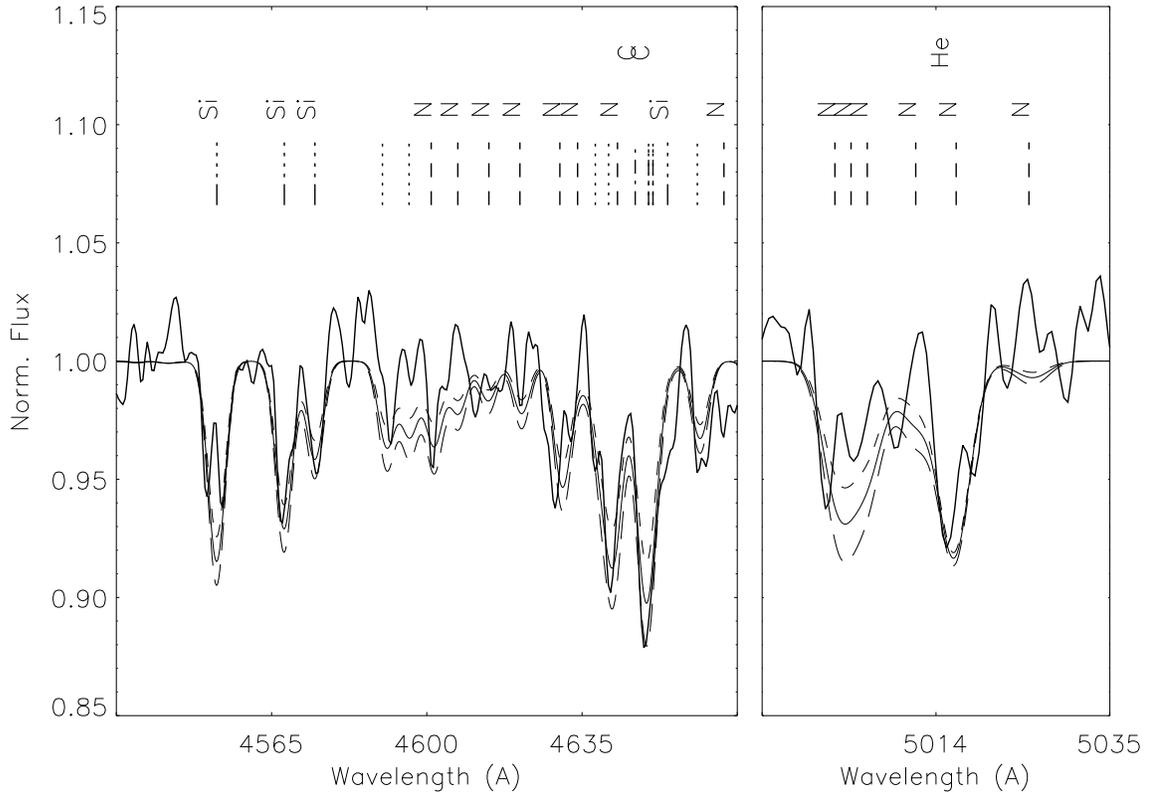}
 \caption{Selected regions of the observed spectrum of A-9, the final fit and two models at $\pm 0.2$ {\it dex} to illustrate
 the uncertainties of the analysis. Line identification as in the previous figure.\label{fig4}}
\end{figure}

Best-fitting models are displayed in Fig. \ref{fig1} and the results summarized in Tab.
\ref{tabla1}. The derived $\beta$ values are consistent with those obtained by \citet[][]{kudritzki1999}
for Galactic B-Sg, with lower values excluded by the arise of emission wings in the synthetic
H$\alpha$ profiles. We estimated an uncertainty of $\pm$0.25 in $\beta$.
In the case of B-12 only the higher Balmer lines have been
considered in the surface gravity determination, as the cores of H$\gamma$ and H$\beta$ are particularly affected by the sky substraction. 
As it has been quoted, the O and N
abundances are very well constrained because of the large number of features from these
species. The presence of a lot of weak metal lines in the 4600 - 4700 \AA~wavelength 
range makes the selection of the continuum level in this area difficult, good
S/N ratio is also needed to disentangle between a real feature and the noise effects.
Final abundance 
uncertainties are estimated to be $\pm\ 0.2$ {\it dex} from model comparisons (see Fig. \ref{fig4}).

We define the mean metallicity as the sum of the $\alpha$-elements abundances, $X_{Si} + 
X_{Mg} + X_{O}$ and refer it to the Sun abundances by \citet[][]{grevesse1998}; at the early stages of 
massive star evolution, 
the O surface abundance 
is not affected by the CNO cycle, which means that the abundance of the $\alpha$-elements is a direct 
measurement of the ZAMS metallicity of the star. The results for B-12, located close to the
galactic center, resembles the abundance patterns of the early B-Sg
in the solar neighborhood, having a solar metallicity within the uncertainties of 
the analysis. 
On the other hand A-9, in the outskirts of the galaxy, has clearly a lower metallicity, around 0.3 $Z_\odot$.
This is in agreement with the results for A-8, a B9-A0 supergiant close to A-9, by 
\citet[][see the Fig. 2]{bresolin2002a}. These authors find a mean metallicity of 0.2 $Z_\odot$ for A-8. We 
must emphasize that both the model atmospheres and the metallicity indicators are different, but the 
results agree extremely well. The metallicity and the spatial location of both stars in NGC 300 points to a
M33-like radial metallicity gradient. The CNO abundances indicate a different degree of chemical evolution, 
while B-12 displays a normal CNO spectrum, A-9 shows indications of strong N enrichment.

\begin{figure}
 \plotone{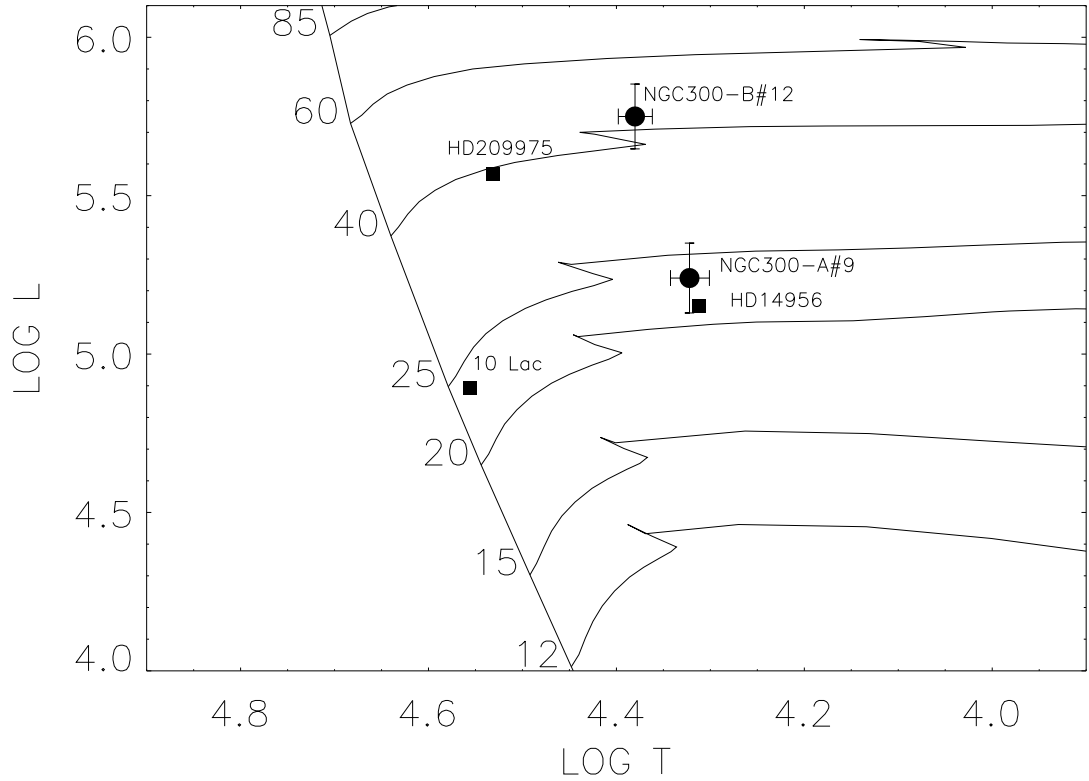}
 \caption{Location of the NGC 300 B-type supergiants in the HR diagram. 
 We have also added the Galactic stars HD14956, HD209975
 and 10 Lac. The theoretical stellar tracks without rotation at solar metallicity are taken 
 from \citet[][]{schaller1992}. Note that A-9 has a metallicity of only 0.3$Z_\odot$.\label{fig2}}
\end{figure}

Synthetic magnitudes and colors (see Tab. \ref{tabla3}) are consistent with almost no reddening for both stars,
except the observed $(V-I)$ for B-12 that seems to be anomalous, probably reflecting the presence of the 
\ion{H}{2} region.
Fig. \ref{fig2} shows the location of the stars on the Hertzprung-Russel diagram, along with theoretical stellar
tracks without rotation at solar metallicity from \citet[][]{schaller1992}. We have also added the location of
the Galactic stars 10 Lac, HD209975 and HD14956 as a reference. 

\begin{figure}
 \plotone{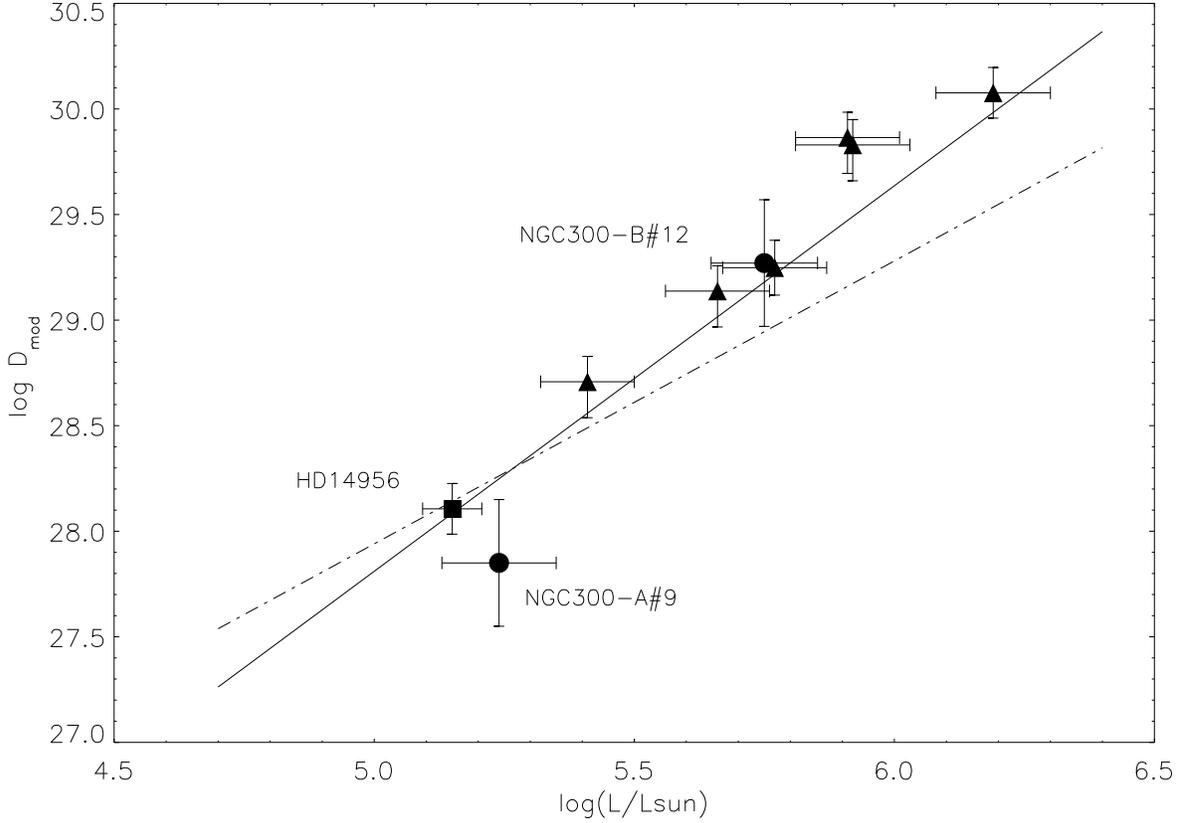}
 \caption{Wind momentum - Luminosity Relationship. Filled circles stand for the NGC
 300 stars and the square denotes HD14956, while the triangles represent the Cyg OB2 
 OB supergiants from Herrero et al. (2002). Solid line is the theoretical prediction by 
 \citet[][]{vink2000} for $T_{eff} \geq 27500\ K$, and the dashed-dotted line is the fit to the
 Galactic early B-type supergiant data from \citet[][]{kudritzki1999}. The modified wind momentun 
 is defined as $D_{\rm mod}\ =\ \dot{M} v_\infty (R_*/R_\odot)^{1/2}$. \label{fig5}}
\end{figure}

Comparing the wind momentum of both NGC 300 stars with the results for Galactic supergiants (Fig.
\ref{fig5}), 
B-12 agrees well with the results by Herrero et al. (2002) for O-type supergiants in the Galactic association
Cyg OB2, as does HD14956. Note, however that the Herrero et al. (2002) stars are considerably hotter than the
ones considered here. The wind momentum of A-9 is also compatible with the WLR of Galactic
early B-Sg as derived by \citet[][]{kudritzki1999}. With respect to this relationship, however, 
B-12 (being an early B-type supergiant as well) shows an enhanced wind momentum rate, which
might be related to clumping effects in the wind that would produce an overestimation of the
mass-loss rate. The failure of our models to reproduce the blue absorption of H$\alpha$ for B-12, in parallel with
an H$\gamma$ core which is too strongly refilled 
might then be explained by this effect, at least
in part, and not only by the rather problematic sky substraction outlined above.
The location of A-9, compared to HD14956, reflects the lower metal content of the NGC 300
supergiant. It must be considered here that we have adopted the same $v_\infty$ for both stars,
HD14956 and A-9, while a lower
value for A-9 could be expected due to its lower metallicity \citep[][]{kudritzki2000}. In any case the effect of 
the lower wind terminal
velocity would reduce even more the wind momentum of A-9, reinforcing the difference with respect to the Galactic
B1.5Ia.

Recently \citet[][]{kudritzki2003} have proposed a new extragalactic distance indicator, the "Flux-weighted -
Luminosity Relationship (FGLR)". The results for both NGC 300 B-type supergiants, B-12 and A-9, follow this 
relationship, within the observed scatter (see the Fig. 2 of the latter reference).

\acknowledgements

We are gratefull to L. J. Corral for making us available the spectrum of HD14956. MAU thanks F. Najarro for providing 
the routines for the computation of the synthetic magnitudes. AH and MAU thank the Spanish MCyT for a support under 
proyect PNAYA2001-0436, partially funded with FEDER funds from the EU. WG gratefully acknowledges financial support 
for this work from the Chilean Center for Astrophysics FONDAP 15010003.

\clearpage

\begin{deluxetable}{lccccccccccccccrc}
 \rotate
 \tabletypesize{\scriptsize}
 \tablecaption{Best-fit model parameters \label{tabla1}}
 \tablewidth{0pt}
 \tablehead{
 \colhead{ID} & \colhead{Teff}   & \colhead{log g}   &
 \colhead{R} &
 \colhead{$Y_{He}$}  & \colhead{$v_{turb}$} & \colhead{$v_\infty$} &
 \colhead{$\dot{M}$} & \colhead{$\beta$}  &
 \colhead{$\epsilon_{Si}$} &
 \colhead{$\epsilon_{O}$}  &
 \colhead{$\epsilon_{Mg}$} &
 \colhead{$\epsilon_{N}$}  &
 \colhead{$\epsilon_{C}$}  &
 \colhead{$Z$} 	           &
 \colhead{$[M/H]$}         &
 \colhead{$log(L/L_\odot)$}\\
 \colhead{ } & \colhead{{\it kK}}   & \colhead{{\it dex}}   &
 \colhead{$R_\odot$} &
 \colhead{ }  & \colhead{$km\ s^{-1}$} & \colhead{$km\ s^{-1}$} &
 \colhead{$10^{-6}\ M_\odot yr^{-1}$} & \colhead{ }  &
 \colhead{{\it dex}} &
 \colhead{{\it dex}}  &
 \colhead{{\it dex}} &
 \colhead{{\it dex}}  &
 \colhead{{\it dex}}  &
 \colhead{$Z_\odot$}  &
 \colhead{dex} 	      &
 \colhead{{\it cgs}}   	
 }
 \startdata
B-12 & 24.0$\pm$1.0 & 2.60$\pm$0.15 & 43.5$\pm$1.5 & 0.10 & 20. & 1500. & 3.00$\pm$0.50 & 1.50 & 7.45 & 8.65 & 7.50 & 7.50 
  & 8.00 & 1.00 &  0.00 & 5.75$\pm$0.10 \\
A-9  & 21.0$\pm$1.0 & 2.50$\pm$0.15 & 32.0$\pm$1.0 & 0.10 & 15. &  800. & 0.25$\pm$0.07 & 2.00 & 7.10 & 8.30 & 7.20 & 8.00 
  & 7.60 & 0.30 & -0.50 & 5.24$\pm$0.11 \\
 \enddata
\tablecomments{The metal abundances are expressed as $\epsilon_A=12+log(A/H)$, while
$[M/H]= (M/H)_* - (M/H)_\odot$.} 
\end{deluxetable}

\clearpage

\begin{deluxetable}{lccccccccccc}
 \tabletypesize{\scriptsize}
 \tablecaption{Observed and synthetic magnitudes and colors.\label{tabla3}}
 \tablewidth{0pt}
 \tablehead{
 \colhead{} & \multicolumn{3}{c}{Observed} & \colhead{} & \multicolumn{4}{c}{Synthetic} & 
 \colhead{} & \colhead{} & \colhead{} \\ 
 \cline{2-4} \cline{6-9} \\
 \colhead{ID}    &  
 \colhead{$V$}   & \colhead{$B-V$}     & \colhead{$V-I$} & \colhead{} &
 \colhead{$M_V$} & \colhead{$M_B-M_V$} & \colhead{$M_V-M_I$} & \colhead{$BC$} & 
 \colhead{} & \colhead{$E(B-V)$} & \colhead{$E(V-I)$}  
 }
 \startdata
B-12 & 19.30 & -0.18 & 0.00 & & -7.29 & -0.17 & -0.23  & -2.33 & & 0.00 & 0.23  \\
A-9  & 20.23 & -0.17 &\ldots & & -6.36 & -0.16 & -0.20     & -1.97 & & 0.00 &\ldots \\
 \enddata

 \tablecomments{We have adopted a distance modulus $\mu\ =\ 26.53\ \pm\ 0.07$ 
 \citep[][]{freedman2001}.}

\end{deluxetable}


\clearpage

\clearpage

\clearpage

\clearpage

\begin{thebibliography}{}

\bibitem[Bresolin et al.(2001)]{bresolin2001} Bresolin, F., Kudritzki, R.-P., M\'endez, R. H., \&
 Przybilla, N. 2001, \apj, 548, 149

\bibitem[Bresolin et al.(2002a)]{bresolin2002a} Bresolin, F., Gieren, W., Kudritzki, R.-P., 
Pietrzy\'nki, G., \& Przybilla, N. 2002a, \apj, 567, 277

\bibitem[Bresolin et al.(2002b)]{bresolin2002b} Bresolin, F., Kudritzki, R.-P., Najarro, F., Gieren, W.
\& Pietrzy\'nki, G. 2002b, \apj, 577, L107

\bibitem[Freedman et al.(2001)]{freedman2001} Freedman, W. L., et al. 2001, \apj, 553, 47
	
\bibitem[Grevesse \& Sauval(1998)]{grevesse1998} Grevesse, N. \& Sauval, A. J. 1998, Space Sci. Rev., 85, 161
	
\bibitem[Haser(1995)]{haser1995} Haser, S. M. 1995, Ph.D. thesis, Ludwing-Maximillians Univ., Munich

\bibitem[Herrero, Puls \& Najarro(2002)]{herrero2002} Herrero, A., Puls, J., \& Najarro, F. 2002, \aap,
in press

\bibitem[Hubeny, Heap \& Lanz(1998)]{hubeny1998} Hubeny, I., Heap, S. R., \& Lanz, T. 1998, ASP Conf. Series Vol
131, 108

\bibitem[Kewley \& Dopita(2002)]{kewley2002} Kewley, L. J. \& Dopita, M. A. 2002, 
\apjs, 142, 35



\bibitem[Kudritzki et al.(1999)]{kudritzki1999} Kudritzki, R.-P., et al. 1999, \aap, 350, 970

\bibitem[Kudritzki \& Puls(2000)]{kudritzki2000} Kudritzki, R.-P. \& Puls, J. 2000, 
\araa, 38, 613

\bibitem[Kudritzki, Bresolin \& Przybilla(2003)]{kudritzki2003} Kudritzki, R.-P., Bresolin, 
F., \& Przybilla, N. 2003, \apj, 582, 83

\bibitem[Monteverde, Herrero, \& Lennon(2000)]{monteverde2000} Monteverde, M. I., Herrero, A., \& Lennon, D. J. 
 2000, \aap, 545, 813

\bibitem[Pilyugin(2002)]{pilyugin2002} Pilyugin, L. S. 2002, preprint (astro-ph/0211319)

\bibitem[Przybilla(2002)]{przybilla2002} Przybilla, N. 2002, Ph.D. thesis, Ludwing-Maximillians Univ., Munich

\bibitem[Puls(1991)]{puls1991} Puls, J. 1991, \aap, 248, 581

\bibitem[Santolaya-Rey, Puls, \& Herrero(1997)]{santolayarey1997} Santolaya-Rey, E., Puls, J., \& Herrero,
A. 1997, \aap, 323, 488

\bibitem[Schaller et al.(1992)]{schaller1992} Schaller, G., Schaerer, D., Meynet, G., \& Maeder, G. 1992, 
\aaps, 96, 269

\bibitem[Smartt, Dufton, \& Lennon(1997)]{smartt1997} Smartt, S. J., Dufton, P. L., \& Lennon, D. J. 1997,
\aap, 326, 763

\bibitem[Smartt et al.(2002)]{smartt2002} Smartt, S. J., Lennon, D. J., Kudrityzki, R.-P., Rosales,
F., Ryans, R. S. I., \& Wright, N. 2002, \aap, 979, 991

\bibitem[Takeda \& Takada-Hidei(1998)]{takeda1998} Takeda, Y. \& Takada-Hidai, M. 1998, \pasj, 50, 629

\bibitem[Venn(1995)]{venn1995} Venn, K. A. 1995, \apj, 449, 839

\bibitem[Villamariz et al.(2002)]{villamariz2002} Villamariz, M. R., Herrero,
A., Becker, S. R., \& Butler, K. 2002, \aap, 388, 940

\bibitem[Vink, de Koter, \& Lamers(2000)]{vink2000} Vink, J. S., de Koter,
A., \& Lamers, H. J. G. L. M. 2000, \aap, 362, 295

\end{thebibliography}
\end{document}